\documentclass[journal]{IEEEtran}

\usepackage{graphicx}
\usepackage{epsfig}
\usepackage{dcolumn}
\usepackage{bm}
\usepackage{amsmath}
\usepackage{amssymb}

\begin{document}
%
\title{Emulation of floating memcapacitors and meminductors using current conveyors}
%
%
%

\author{Yuriy~V.~Pershin and Massimiliano~Di~Ventra
\thanks{Yu. V. Pershin is with the Department of Physics
and Astronomy and USC Nanocenter, University of South Carolina,
Columbia, SC, 29208 \newline e-mail: pershin@physics.sc.edu.}
\thanks{M. Di Ventra is with the Department
of Physics, University of California, San Diego, La Jolla,
California 92093-0319 \newline e-mail: diventra@physics.ucsd.edu.}

\thanks{Manuscript received November XX, 2010; revised January YY, 2011.}}

%
%


\maketitle

\begin{abstract}
We suggest circuit realizations of emulators transforming memristive devices into
effective floating memcapacitive and meminductive systems. The emulator's circuits
are based on second generation current conveyors and involve either four single-output or two dual-output
current conveyors. The equations governing the resulting memcapactive and meminductive systems are presented.
\end{abstract}


%
\IEEEpeerreviewmaketitle

\IEEEPARstart{T}{he} class of memory circuit elements \cite{diventra09a} involves memristors \cite{chua71a,chua76a},
memcapacitors \cite{diventra09a} and meminductors \cite{diventra09a}, namely resistors, capacitors and inductors, respectively, which retain memory of the past dynamics.
While there are many discovered experimental realizations of memristive systems, the number of systems showing memcapacitive and meminductive behavior is still very limited \cite{diventra09a}. Therefore, electronic circuits that emulate the behavior of memcapacitive and meminductive elements are of significant interest. One way to obtain memcapacitive or meminductive response is to use mutators \cite{pershin10b,biolek10b}. However, recently suggested mutators have some limitations~\cite{pershin10b,biolek10b}. For instance, the mutators discussed in reference \cite{pershin10b} realize grounded memcapacitive and meminductive systems connected
in series with a resistor. The mutator from reference \cite{biolek10b} simulates a grounded memcapacitive system. For applications in electronics, however, it is desirable to have emulators of memcapacitive and meminductive systems that can be connected between any two voltages, and that do not involve any additional resistances. Such circuits are proposed in this work.

The suggested circuits transform a time-dependent response of a memristive system into effective floating memcapacitive and meminductive responses. The memristive system is defined by the following equations:
\begin{eqnarray}
I_M&=&\left[ R_M\left( x,V_M,t \right)\right]^{-1}V_M, \label{RM1}  \\
\frac{\textnormal{d}x}{\textnormal{d}t}&=&f\left( x,V_M,t \right), \label{RM2}
\end{eqnarray}
where $I_M$ and $V_M$ are the current through and the voltage across the memristive system, respectively \cite{chua76a}, $R_M$ is the memristance (time-dependent resistance), $x$ is a vector of the internal state variables and $f$ is a device-specific function. The memristive system could be realized by a physical material/system \cite{diventra09a} (with $f$ given, e.g., by the equations in Ref. \cite{pershin09b})
or it could be emulated as suggested by the present authors in Refs. \cite{pershin10b,pershin09d,pershin09c}.

\begin{figure}
 \begin{center}
\includegraphics[width=6cm]{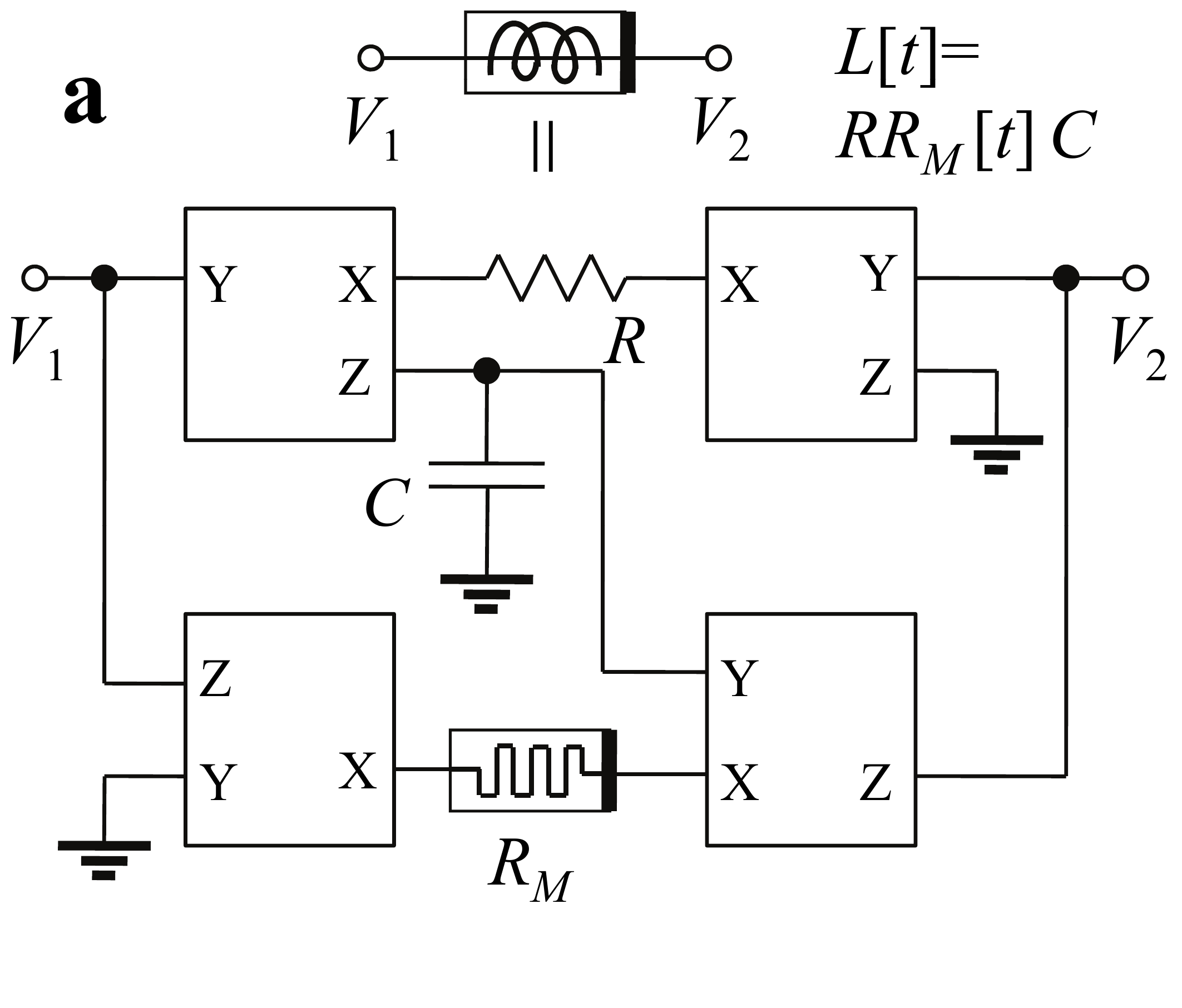}
\vspace{0.3cm}
\includegraphics[width=6cm]{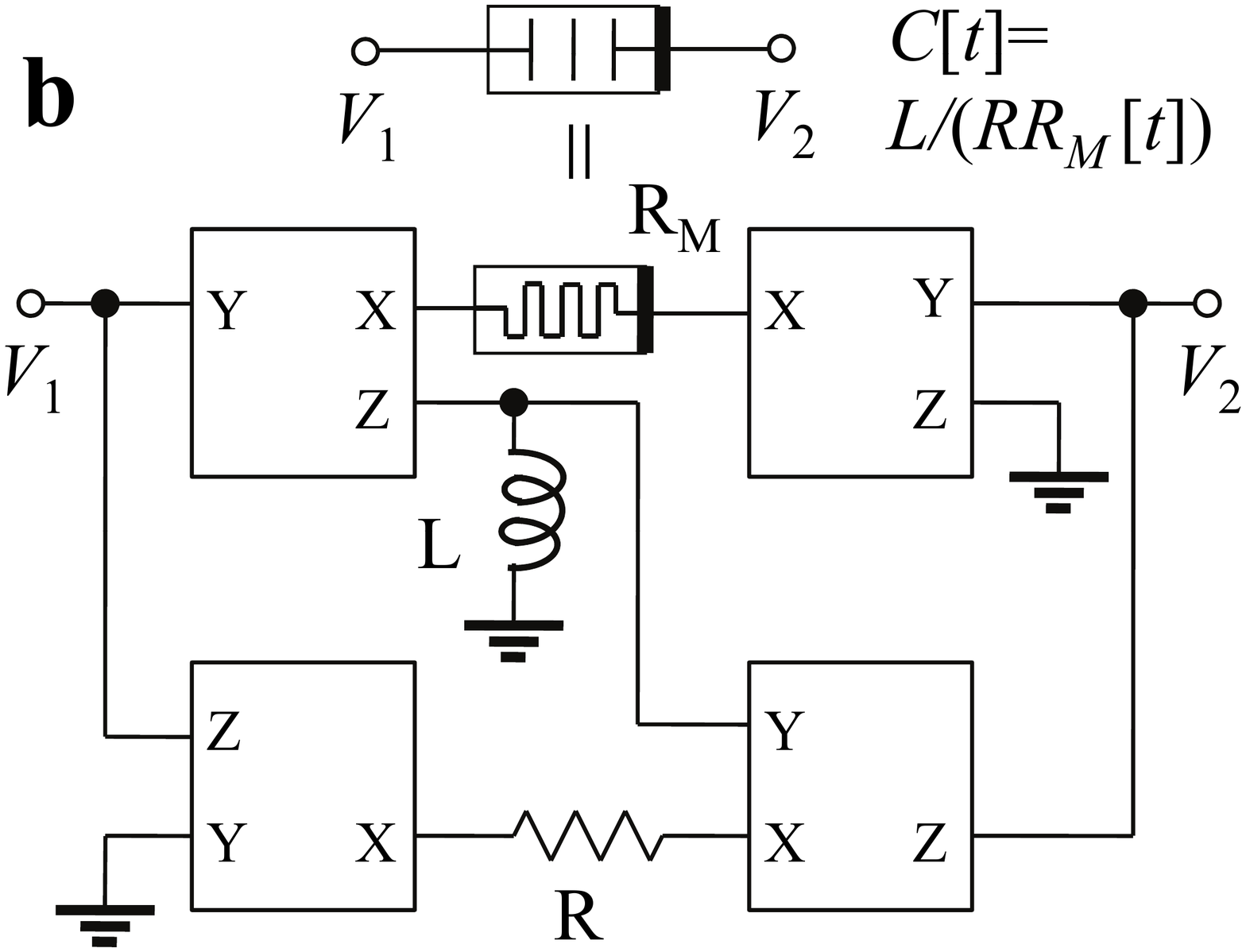}
\caption{Emulators of meminductive (a) and memcapacitive (b) systems employing single-output current conveyors. Here, $[t]$ denotes implicit and explicit time dependence of corresponding elements. \label{fig1}}
 \end{center}
\end{figure}

Figure \ref{fig1} shows circuits of mutators employing four single-output second generation current conveyors (CCII+) \cite{sedra70a,wilson90a}. The main circuit's topology is based on the floating inductance simulation circuit \cite{kiranon97a}. In addition to four active elements, each circuit uses three passive elements such as the memristive system $R_M$ described by Eqs. (\ref{RM1}-\ref{RM2}), and standard linear $R$, $C$ and $L$. The operation of a current conveyor can be characterized by the following relations: $V_x=V_y$, $I_x=I_z$, $I_y=0$. By straightforward analysis of the circuit shown in Fig. \ref{fig1}(a) we find that the circuit's response between terminals 1 and 2 is described by
\begin{eqnarray}
I&=&\frac{1}{C R R_M\left( x,\phi/(RC),t \right)}\phi \equiv \left[ L(x,\phi,t) \right]^{-1}\phi, \label{L1} \\
\frac{\textnormal{d}x}{\textnormal{d}t}&=&f\left( x,\phi/(RC) ,t \right)  \label{L2}
\end{eqnarray}
where the flux $\phi$ is given by $\phi(t)=\int\limits_0^t\left(V_1(t')-V_2(t') \right)dt'$. The analysis of the circuit presented in Fig. \ref{fig1}(b) results in the following system's equations:
\begin{eqnarray}
Q&=&\frac{L}{R R_M\left( x,(V_1-V_2),t \right)}\left(V_1-V_2 \right) \equiv  \nonumber \\
&\equiv&C\left( x,(V_1-V_2),t \right) \left( V_1-V_2\right), \label{C1} \\
\frac{\textnormal{d}x}{\textnormal{d}t}&=&f\left( x,(V_1-V_2) ,t \right) . \label{C2}
\end{eqnarray}
It is readily seen that Eqs. (\ref{L1}-\ref{L2}) and (\ref{C1}-\ref{C2}) describe {\it floating} flux-controlled meminductive systems \cite{diventra09a} and voltage-controlled memcapacitive systems \cite{diventra09a}, respectively.
If we consider equation (\ref{L1}) and algebraically solve it
with respect to the flux $\phi$ (assuming a unique solution at any given time), we can then substitute the result into~(\ref{L2}) and obtain equations of a current-controlled meminductive
system. The same can be done for Eqs.~(\ref{C1}) and~(\ref{C2}) thus transforming a voltage-controlled memcapacitive system into a charge-controlled one, provided a unique solution of Eq.~(\ref{C1}) exists at any given time.

\begin{figure}
 \begin{center}
\includegraphics[width=6cm]{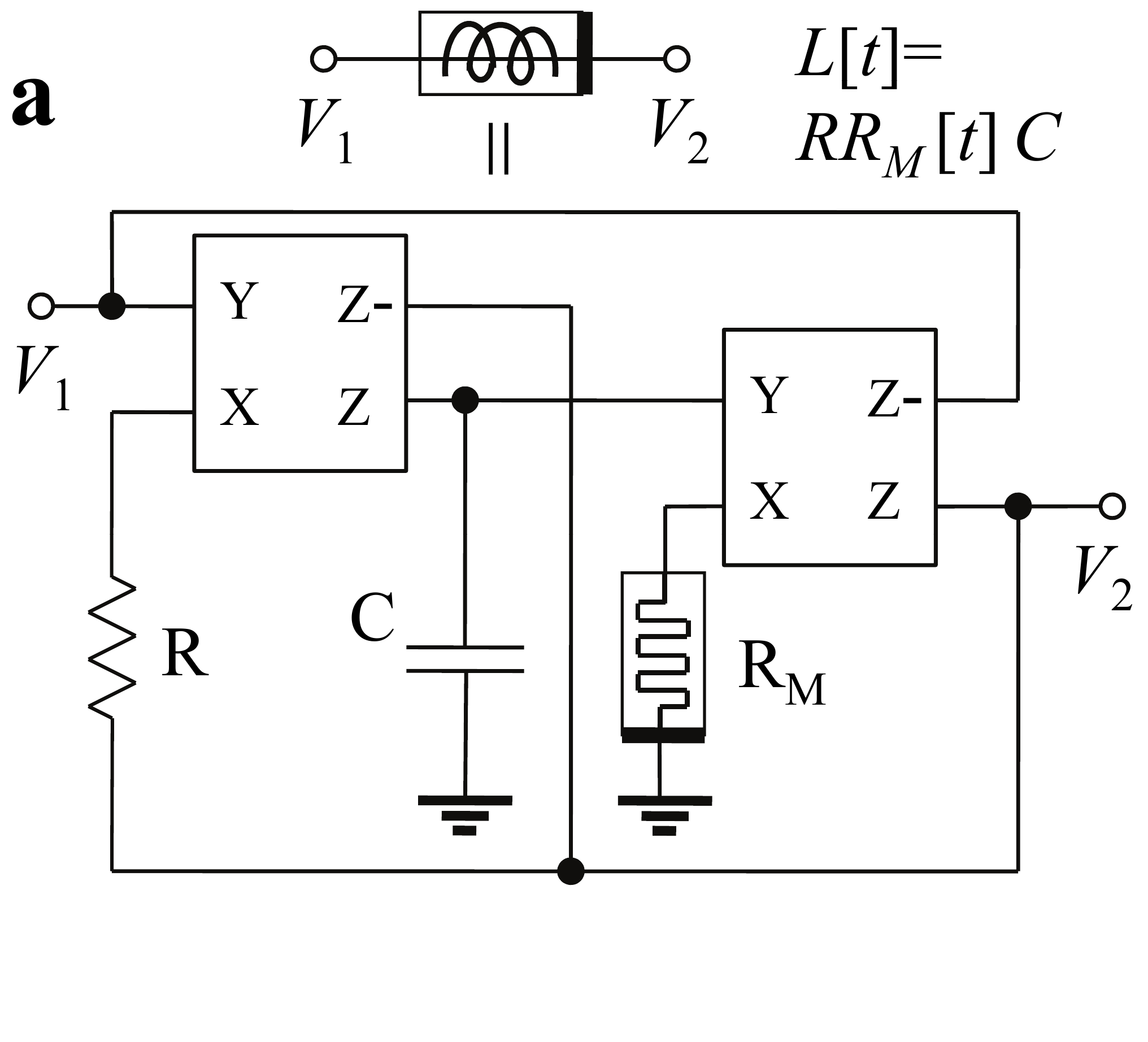}
\vspace{0.3cm}
\includegraphics[width=6cm]{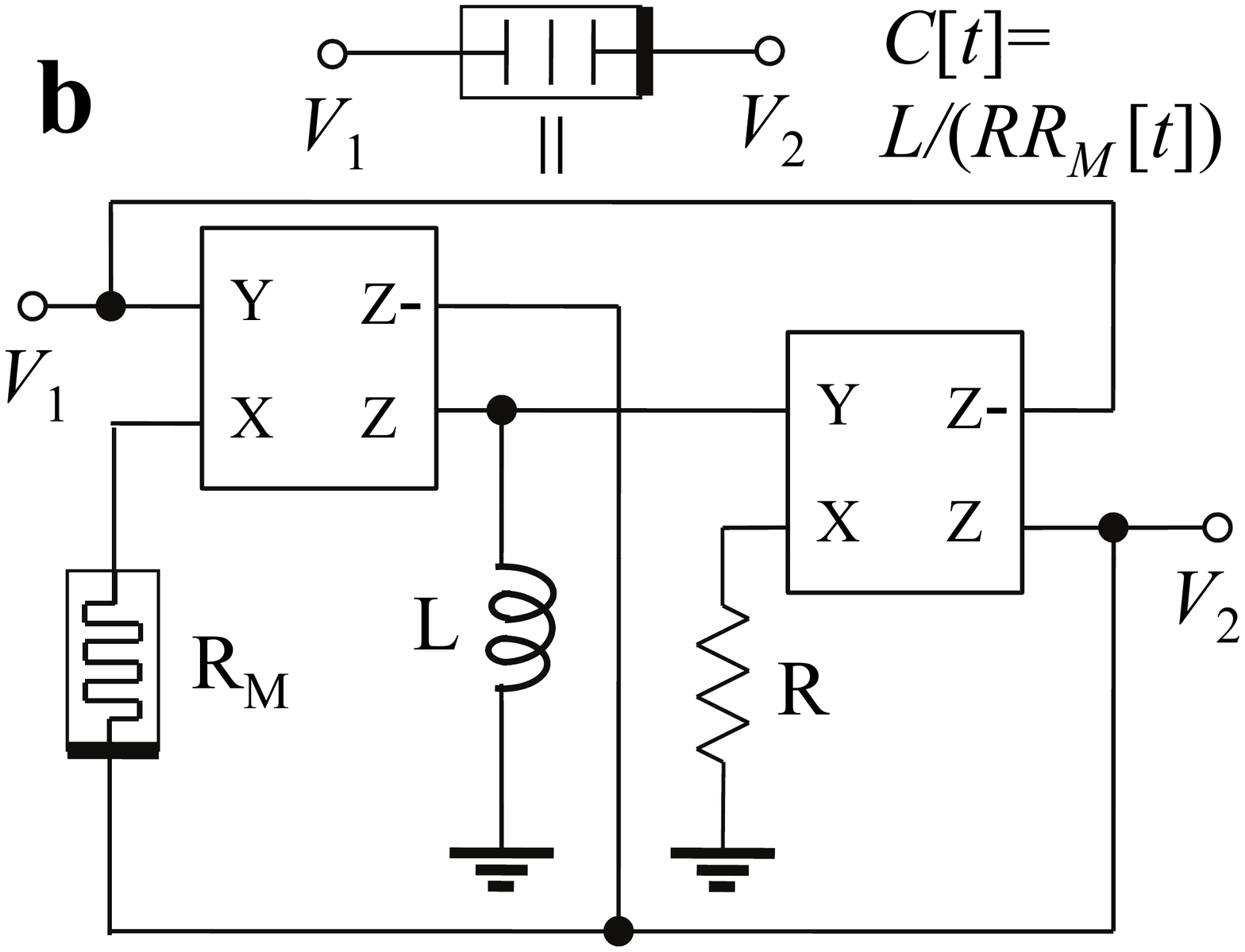}
\caption{Emulators of meminductive (a) and memcapacitive (b) systems employing dual-output current conveyors. The dual-output current conveyors satisfy the rules: $V_x=V_y$, $I_x=I_z$, $I_y=0$, and $I_{z-}=-I_z$. \label{fig2}}
 \end{center}
\end{figure}

It was previously demonstrated that dual-output current conveyors can be used to reduce the number of active components in different circuits \cite{pal89a,Soliman98a,Tangsrirat07a}. Fig. \ref{fig2} represents emulators of meminductive and memcapacitive systems based on a floating-inductor simulation circuit \cite{pal89a}. Each of the  circuits in fig. \ref{fig2} employs one memory element (memristor), two other passive and two active elements. The equations governing the effective meminductive and memcapacitive behaviors are the same as Eqs. (\ref{L1}-\ref{L2}) and Eqs. (\ref{C1}-\ref{C2}), respectively.

In conclusion, we have presented emulators of floating memcapacitive and meminductive systems using second generation current conveyors involving either four single-output or two dual-output
current conveyors. The resulting circuits mutate memristive system's dynamics into an effective memcapacitive and meminductive response. Since memristor emulators can be easily made from off-the-shelf
components \cite{pershin09d,pershin09c,pershin10b}, we expect these mutators to find widespread use in electronic applications with memdevices.

\bibliographystyle{IEEEtran}
\bibliography{IEEEabrv,memcapacitor}

\end{document}